\documentstyle[prl,aps,floats,epsf]{revtex}

\begin{document}
\draft

\twocolumn[\hsize\textwidth\columnwidth\hsize\csname
@twocolumnfalse\endcsname

\title{
Formation and rapid evolution of  
domain structure at phase transitions 
in slightly inhomogeneous  ferroelectrics
}
\author{A.M. Bratkovsky$^{1}$ and A.P. Levanyuk$^{1,2}$}

\address{$^{1}$ Hewlett-Packard Laboratories, 1501 Page Mill Road, Palo
Alto, California 94304\\
$^{2}$Departamento de F\'{i}sica de la Materia Condensada, C-III,
Universidad Aut\'{o}noma de Madrid, 28049 Madrid, Spain
}
\date{November 29, 2001 }
\maketitle

\begin{abstract}

We present the first analytical study of stability loss and
evolution of domain structure in inhomogeneous
ferroelectric samples for exactly solvable model.
The model assumes a short-circuited capacitor
 with two regions with slightly
different critical temperatures $T_{c1}>T_{c2}$,
where $T_{c1}-T_{c2} \ll T_{c1}, T_{c2}$. 
Even a tiny inhomogeneity like $10^{-5}$K
may result in splitting the system into domains below the phase
transition temperature.
At $T<T_{c2}$ the domain width $a$  is proportional to 
$(T_{c1}-T)/(T_{c1} - T_{c2})$ and quickly increases with
lowering temperature. 
The minute inhomogeneities in $T_c$ may result from structural 
(growth) inhomogeneities which are always present in real samples and 
a similar role can be played by inevitable temperature gradients.


\end{abstract}
\vskip 2pc ]

The idea that the phase transition in electroded short-circuited
ferroelectric proceeds into homogeneous monodomain state\cite{Ginzburg49} is
very well known. Similar result also applies to free ferroelastic crystals.
However, it has {\em never} been observed. Surprisingly, both electroded
ferroelectrics and free ferroelastics do split into domains, although they
should not. The present paper aims to answer why.

It is generally assumed that in the finite ferroelectric samples the domain
structure appears in order to reduce the depolarizing electric field if
there is a nonzero normal component of the polarization at the surface of
the ferroelectrics \cite{Ginzburg49,Kanzig52} (in complete analogy with
ferromagnets \cite{Landau35}), if the field cannot be reduced by either
conduction (usually negligible in ferroelectrics at low temperatures) or
charge accumulation from environment at the surface \cite{Jona62}. On the
other hand, in inhomogeneous ferroelastics (e.g. films on a substrate, or
inclusions of a new phase in a matrix) the elastic domain structure
accompanies the phase transition in order to minimize the strain energy, as
is well understood in case of martensitic phase transformations \cite
{Khachaturyan} and epitaxial thin films \cite{Roytburd76,BLprl2,BLstloss}.

In search for reasons of domain appearance in otherwise perfect electroded
samples, which is not yet understood, we shall discuss a second order
ferroelectric phase transition in slightly inhomogeneous electroded sample.
This problem has not been studied before. We consider an exactly solvable
case of a system, which has two slightly different phase transition
temperatures in its two parts. While the phase transition occurs in the
``soft'' part of the system, the ``hard'' part may effectively play a role
of a ``dead'' layer \cite{BLprl1} and trigger a formation of the domain
structure in the ``soft'' part with fringe electric fields penetrating the
``hard'' part. One has to check this possibility, but the behavior of the
corresponding domain structure is expected to be unusual: it should strongly
depend on temperature since further cooling transforms the ``hard'' part
into a ``soft'' one, while the first ``soft'' part becomes ``harder''. Since
the inhomogeneity is small, one might expect that the domains would quickly
grow with lowering temperature. We indeed find a rapid growth of the domain
width linearly with temperature in the case of slightly inhomogeneous
short-circuited ferroelectric. This behavior is generic and does not depend
on particular model assumptions. Generally, the inhomogeneous ferroelectric
systems pose various fundamental problems and currently attract a lot of
attention. In particular, {\em graded} ferroelectric films and ferroelectric 
{\em superlattices} have been shown to have giant pyroelectric \cite{graded}
and unusual dielectric response\cite{FEsuper}.

We shall consider the case of slightly inhomogeneous uniaxial ferroelectric
in short-circuited capacitor that consists of two layers with slightly
different critical temperatures, so that, for instance, a top part
``softens'' somewhat earlier than the bottom part does. We assume the easy
axis $z$ perpendicular to electrode plates, and make use of the Landau free
energy functional for given potentials on electrodes (zero in the present
case)\cite{LLvol8} $\tilde{F}=F_{LGD}\left[ \vec{P}\right] +\int dV\left( -%
\vec{E}\vec{P}-\frac{E^{2}}{8\pi }\right) ,$ with 
\begin{eqnarray}
F_{LGD}[\vec{P}] &=&\sum_{p=1,2}\int dV[\frac{A_{p}}{2}P_{z}^{2}+\frac{B}{4}%
P_{z}^{4}  \nonumber \\
&&+\frac{D}{2}\left( \nabla _{\perp }P_{z}\right) ^{2}+\frac{g}{2}\left(
\partial _{z}P_{z}\right) ^{2}+\frac{A_{\perp }}{2}\vec{P}_{\perp }^{2}],
\label{eq:lgd}
\end{eqnarray}
where $P_{z}$ $\left( \vec{P}_{\perp }\right) $is the polarization component
along (perpendicular to) the ``soft'' direction, index $p=1(2)$ marks the
top (bottom) part of the film: 
\begin{eqnarray*}
A_{1} &=&A,\qquad 0<z<l_{1}, \\
A_{2} &=&A+\delta A,\qquad -l_{2}<z<0,
\end{eqnarray*}
where $A_{1(2)}=\alpha (T-T_{c1(2)})$ and $\delta A>0$ (meaning $%
T_{c2}<T_{c1})$. The constant $\alpha =1/T_{0},$ where $T_{0}\sim T_{at}$ $%
(\sim T_{c})$ for displacive (order-disorder)\ type ferroelectrics.

The equation of state is $\delta F_{LGD}[P]/\delta \vec{P}=\vec{E}=-\nabla
\varphi ,$ where $\varphi $ is the electrostatic potential, or in both parts
of the film $p=1,2:$%
\begin{eqnarray}
E_{z} &=&-\partial _{z}\varphi =A_{p}P_{z}+BP_{z}^{3}-D\nabla _{\perp
}^{2}P_{z}-g\partial _{z}^{2}P_{z},  \label{eq:Ez} \\
\vec{E}_{\perp } &=&A_{\perp }\vec{P}_{\perp },  \label{eq:Eperp}
\end{eqnarray}
These equations should be solved together with the Maxwell equation, ${\rm %
div}(\vec{E}+4\pi \vec{P})=0,$ or 
\begin{equation}
\left( \partial _{z}^{2}+\epsilon _{a}\nabla _{\perp }^{2}\right) \varphi
=4\pi \partial _{z}P_{z},  \label{eq:div1}
\end{equation}
where the dielectric constant in the plane of the film is $\epsilon
_{a}=1+4\pi /A_{\perp }.$

{\em Loss of stability}${\em .-}${\em \ }We shall now find conditions for
loss of stability of the paraelectric phase close to $T_{c1}$ with respect
to inhomogeneous polarization. At the point of stability loss the
polarization is small and the nonlinear term $P_{z}^{3}$ must be omitted. We
are looking for a nontrivial solution in a form of the ''polarization
wave'', 
\begin{equation}
P_{z},\varphi \propto e^{ikx}.
\end{equation}
We shall check later that the stability will be lost for the wave vector $%
kl_{1}\gg 1,$ so that $\nabla _{\perp }^{2}P_{z}=k^{2}P_{z}\gg g\partial
_{z}^{2}P_{z}\sim P_{z}/l_{1}^{2},$ and the last term in the right-hand side
of (\ref{eq:Ez})\ should be dropped. Going over to Fourier harmonics
indicated by the subscript $k$, we obtain 
\begin{equation}
\varphi _{k}^{\prime \prime }-\epsilon _{a}k^{2}\varphi _{k}=4\pi
P_{zk}^{\prime },  \label{eq:fiP}
\end{equation}
where the prime indicates derivative $($ $f^{\prime }\equiv df/dz,$ $%
f^{\prime \prime }\equiv d^{2}f/dz^{2})$. We can exclude $P_{zk}$ with the
use of the linearized equation of state (\ref{eq:Ez}), which gives 
\begin{equation}
-\varphi _{k}^{\prime }=(A_{p}+Dk^{2})P_{zk}.  \label{eq:lineq}
\end{equation}
Substituting this into (\ref{eq:fiP}), we obtain $\varphi _{k}^{\prime
\prime }-\frac{\epsilon _{a}k^{2}\left( A_{p}+Dk^{2}\right) }{4\pi }\varphi
_{k}=0,$ where we have used $\left| A+Dk^{2}\right| /4\pi \ll 1,$ which is
always valid in ferroelectrics. We shall see momentarily that the nontrivial
solution appears only when $A_{1}+Dk^{2}<0,$ while $A_{2}+Dk^{2}>0.$ The
resulting system is 
\begin{eqnarray}
\varphi _{1k}^{\prime \prime }+\chi _{1}^{2}k^{2}\varphi _{1k} &=&0,
\label{eq:f1} \\
\varphi _{2k}^{\prime \prime }-\chi _{2}^{2}k^{2}\varphi _{2k} &=&0,
\label{eq:f2}
\end{eqnarray}
where $\chi _{1}^{2}=-\frac{\epsilon _{a}\left( A_{1}+Dk^{2}\right) }{4\pi }%
, $ $\chi _{2}^{2}=\frac{\epsilon _{a}\left( A_{2}+Dk^{2}\right) }{4\pi }.$
The boundary condition reads as 
\begin{equation}
\frac{\varphi _{1k}^{\prime }}{A_{1}+Dk^{2}}=\frac{\varphi _{2k}^{\prime }}{%
A_{2}+Dk^{2}},  \label{eq:bcprime}
\end{equation}
where we have used $\left| A_{1}+Dk^{2}\right| /4\pi \ll 1.$ We obtain\ from
Eqs. (\ref{eq:f1})-(\ref{eq:bcprime}) the condition for a nontrivial
solution 
\begin{equation}
\chi _{1}\tan \chi _{1}kl_{1}=\chi _{2}\tanh \chi _{2}kl_{2},
\label{eq:disp}
\end{equation}
which has a homogeneous solution $k=0$ and the inhomogeneous solution with $%
k=k_{c}$ (\ref{eq:kcel}), hence we have to determine which one is actually
realized. The inhomogeneous solution is easily found for $\chi
_{2}kl_{2}\gtrsim 1,$ where $\tanh $ can be replaced by unity. Close to the
transition $\chi _{2}/\chi _{1}\gg 1,$ and the solution is 
\begin{equation}
\chi _{1}kl_{1}=\frac{\pi }{2}\frac{\chi _{2}kl_{1}}{1+\chi _{2}kl_{1}}%
\approx \frac{\pi }{2},
\end{equation}
when $\chi _{2}kl_{1}\gg 1.$ This gives $|A|=Dk^{2}+\frac{\pi ^{3}}{\epsilon
_{a}k^{2}l_{1}^{2}}.$There is no solution for $\chi _{1}^{2}<0.$ The minimal
value of $A$ for the nontrivial solution (onset of instability) is defined
by 
\begin{eqnarray}
k_{c} &=&\left( \frac{\pi ^{3}}{\epsilon _{a}Dl_{1}^{2}}\right)
^{1/4}\approx \frac{\pi ^{3/4}}{\epsilon _{a}^{1/4}}\frac{1}{\sqrt{%
d_{at}l_{1}}},  \label{eq:kcel} \\
|A|_{c} &=&2Dk_{c}^{2}=\frac{2\pi ^{3/2}D^{1/2}}{\epsilon _{a}^{1/2}l_{1}}%
\approx \frac{2\pi ^{3/2}}{\epsilon _{a}^{1/2}}\frac{d_{at}}{l_{1}},
\label{eq:Acel}
\end{eqnarray}
where we have introduced the ``atomic'' size $d_{at}\sim \sqrt{D}$
comparable to the lattice parameter. We obtain the corresponding tiny shift
in the critical temperature [see estimates below Eq.(\ref{eq:dTc})] $%
T_{c1}-T_{c}\sim T_{0}d_{at}/\epsilon _{a}^{1/2}l_{1}.$ Hence, the system
looses its stability very quickly below the bulk transition temperature. It
is readily checked that the assumptions we used to obtain the solution are
easily satisfied. Indeed, $\chi _{2}kl_{2}\gtrsim 1$ and $\chi _{2}kl_{1}\gg
1$ both correspond to approximately the same condition when $l_{1}\sim
l_{2}:\delta A\gg \frac{4}{\pi ^{1/2}\epsilon _{a}^{1/2}}\frac{d_{at}}{l_{1}}%
,$ meaning that the difference between $T_{c}$ should be larger than the
shift of $T_{c}$.

Now we have to determine when the transition into inhomogeneous state occurs
prior to a loss of stability with respect to a {\em homogeneous}
polarization. The homogeneous loss of stability corresponds to $A=A_{h}$
found from 
\begin{equation}
A_{h}l_{1}+\left( A_{h}+\delta A\right) l_{2}=0.
\end{equation}
For the inhomogeneous state to appear first, there must be $A_{c}>A_{h},$ or 
$\delta A>\frac{\pi ^{3/2}(l_{1}+l_{2})}{\epsilon _{a}^{1/2}l_{1}}\frac{%
d_{at}}{l_{1}}.$ This means that very {\em tiny inhomogeneity} in the sample
is enough to split it into the domain structure, 
\begin{equation}
T_{c1}-T_{c2}=T_{0}\frac{\pi ^{3/2}(l_{1}+l_{2})}{\epsilon _{a}^{1/2}l_{1}}%
\frac{d_{at}}{l_{1}},  \label{eq:dTc}
\end{equation}
which is estimated as $T_{at}\frac{d_{at}}{\epsilon _{a}^{1/2}l_{1}}\lesssim
\epsilon _{a}^{-1/2}(10^{4}-10^{5})10^{-7}K=(10^{-3}-10^{-2}){\rm K}$ for
displacive systems, and $T_{c}\frac{d_{at}}{\epsilon _{a}^{1/2}l_{1}}%
\lesssim \left( 10^{-5}-10^{-4}\right) ${\rm K} for order-disorder systems.
Certainly, such a small temperature and/or compositional inhomogeneity
exists in all usual experiments.

{\em Domain structure at }$T_{c2}<T<T_{c1}$ $(A<0,${\em \ }$A+\delta A>0).-$
After stability loss the resulting ''polarization wave'' quickly develops
into a domain structure, as we shall now demonstrate. In the region well
below $T_{c1}$ we can use the linearized equation of state 
\begin{equation}
E_{z}=(A+3BP_{01}^{2})(P_{z}-P_{01})=-2A(P_{z}-P_{01}),
\end{equation}
where $\left| P_{01}\right| =\sqrt{-A/B}$ is the spontaneous polarization in
the top layer, which gives $P_{z1}=P_{01}+\frac{1}{2|A|}E_{z}$, $P_{z2}=%
\frac{1}{A_{2}}E_{z},$ for the top and bottom layers, respectively. In this
case the equation for the potential $\varphi $ (\ref{eq:div1}) reduces to a
standard Laplace equation $\left( \epsilon _{c}\partial _{z}^{2}+\epsilon
_{a}\nabla _{\perp }^{2}\right) \varphi =0,$ with the boundary condition 
\begin{equation}
\epsilon _{c1}\partial _{z}\varphi _{1}-\epsilon _{c2}\partial _{z}\varphi
_{2}=4\pi P_{01}(x),
\end{equation}
where $\epsilon _{c1}=1+2\pi /|A|,$ $\epsilon _{c2}=1+4\pi /A_{2}.$

\begin{figure}[t]
\epsfxsize=2.8in \epsffile{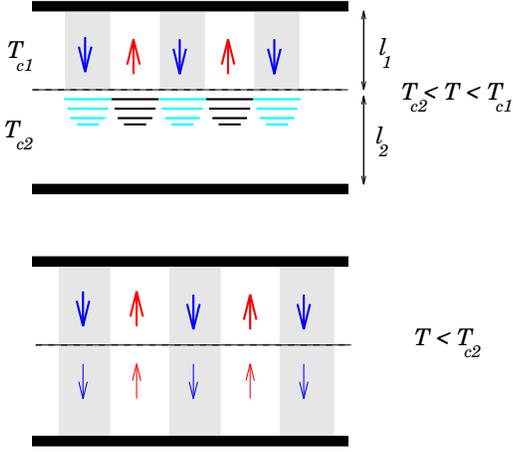}
\caption{ Schematic of the domain structure with the period $2a$ in
inhomogeneous ferroelectric film of the thickness $l_{1}+l_{2}$. Top and
bottom layers have slightly different critical temperatures $T_{c1}>T_{c2}$, 
$T_{c1}-T_{c2}\ll T_{c1},T_{c2}$. Slightly below $T_{c1}$ the top layer
splits into domains with electric fringe field propagating into the bottom
layer (fringe field shown as the hatched area in the top panel). The domains
persist and evolve below $T_{c2}$ when both layers exhibit a ferroelectric
(or ferroelastic) transition (bottom panel). }
\label{fig:fringe}
\end{figure}

The spontaneous polarization in the top layer alternates from domain to
domain as $P_{01}(x)=\pm \left| P_{01}\right| \equiv \pm \sqrt{-A/B}.$ We
are looking for a solution in a form of a domain structure with a period $%
T=2a$ (Fig. \ref{fig:fringe}), 
\begin{equation}
P_{01}(x)=\sum_{k}P_{01k}e^{ikx},\hspace{0.2in}\varphi (x)=\sum_{k}\varphi
_{k}e^{ikx},  \label{eq:fik}
\end{equation}
with $k\equiv k_{n}=2\pi n/T=\pi n/a,$ $n=\pm 1,\pm 2,...$ Going over to the
Fourier harmonics, we can write the Laplace equations for both parts of the
film as 
\begin{eqnarray}
\epsilon _{c1}\varphi _{1k}^{\prime \prime }-\epsilon _{a}k^{2}\varphi _{1k}
&=&0,  \label{eq:lap1} \\
\epsilon _{c2}\varphi _{2k}^{\prime \prime }-\epsilon _{a}k^{2}\varphi _{2k}
&=&0,  \label{eq:lap2}
\end{eqnarray}
with the boundary conditions at the interface $z=0$ 
\begin{equation}
\varphi _{1k}=\varphi _{2k},\qquad \epsilon _{c1}\varphi _{1k}^{\prime
}-\epsilon _{c2}\varphi _{2k}^{\prime }=4\pi P_{01k}  \label{eq:bc01}
\end{equation}
The corresponding {\em electrostatic} (stray) field part of the energy is
found as \cite{BLprl1} $\tilde{F}_{es}=\frac{1}{2}\int d{\cal A}\sigma
_{s}\varphi \left( z=0\right) ,$ where $\sigma _{s}$ is the density of bound
charge at the interface , corresponding to {\em only} the spontaneous part
of the polarization $P_{01}(x),$ and integration goes over the area ${\cal A}
$\ between two parts of the film. We calculate this expression by going over
to Fourier expansion (\ref{eq:fik}) and using the fact that in the present
geometry $\sigma _{s}(x)=-P_{01}(x)$ (and, therefore, its Fourier component $%
\sigma _{sk}=-P_{01k}),$%
\begin{equation}
\frac{\tilde{F}_{es}}{{\cal A}}=\sum_{k>0}\frac{4\pi |P_{01k}|^{2}}{kD_{k}},
\label{eq:Fes01}
\end{equation}
$D_{k}=\epsilon _{a}^{1/2}\left[ \epsilon _{c1}^{1/2}\coth \sqrt{\frac{%
\epsilon _{a}}{\epsilon _{c1}}}kl_{1}+\epsilon _{c2}^{1/2}\coth \sqrt{\frac{%
\epsilon _{a}}{\epsilon _{c2}}}kl_{2}\right] ,$ with $k=\pi n/a,$ $%
n=1,2,..., $ similar to \cite{BLef}. Note that here $P_{01k}=2\left|
P_{01}\right| /i\pi n,$ $n=2j+1,$ $j=0,1,...$ and zero otherwise. Adding the
surface energy of the domain walls, we obtain the free energy of the domain
pattern 
\begin{equation}
\frac{\tilde{F}}{{\cal A}}=\frac{\gamma _{1}l_{1}}{a}+\frac{16P_{01}^{2}a}{%
\pi ^{2}}\sum_{j=0}^{\infty }\frac{1}{\left( 2j+1\right) ^{3}D_{2j+1}},
\label{eq:Ft01}
\end{equation}
where $D_{n}=D_{k_{n}}.$ Not very close to $T_{c}$ the argument of $\coth $
is $\sqrt{\frac{\epsilon _{a}}{\epsilon _{c1}}}kl_{1}\gtrsim 1,$ so that $%
D_{k}=\epsilon _{a}^{1/2}\left( \epsilon _{c1}^{1/2}+\epsilon
_{c2}^{1/2}\right) .$ Minimizing the free energy, we find the domain width 
\begin{equation}
a=\left[ \frac{\pi ^{2}\epsilon _{a}^{1/2}\left( \epsilon
_{c1}^{1/2}+\epsilon _{c2}^{1/2}\right) }{14\zeta (3)}\Delta _{1}l_{1}\right]
^{1/2},  \label{eq:a1}
\end{equation}
where $\Delta _{1}\equiv \gamma _{1}/P_{01}^{2}=d_{at}|A|^{1/2}$ is the
characteristic microscopic length, and $d_{at}\equiv \frac{2^{3/2}}{3}%
D^{1/2} $ is comparable to\ a lattice spacing (``atomic'' length scale)$.$
The expression (\ref{eq:a1}) is valid when $\sqrt{\frac{\epsilon _{a}}{%
\epsilon _{c1}}}kl_{1}\gtrsim 1,$ or $|A|\gtrsim 2d_{at}/\left( \pi \epsilon
_{a}^{1/2}l_{1}\right) ,$ meaning that one has to be below $T_{c}$ by a tiny
amount $T_{c1}-T\gtrsim T_{0}d_{at}/\left( \epsilon _{a}^{1/2}l_{1}\right) ,$
estimated earlier. Note that close to $T_{c1}$ one obtains for the domain
width 
\begin{equation}
a=a_{K}\equiv \left[ \frac{\pi ^{5/2}\epsilon _{a}^{1/2}}{7\sqrt{2}\zeta (3)}%
d_{at}l_{1}\right] ^{1/2},  \label{eq:aKel}
\end{equation}
and this value does {\em not} depend on temperature. We shall formally refer
to this result as the Kittel domain width.

\begin{figure}[t]
\epsfxsize=2.4in \epsffile{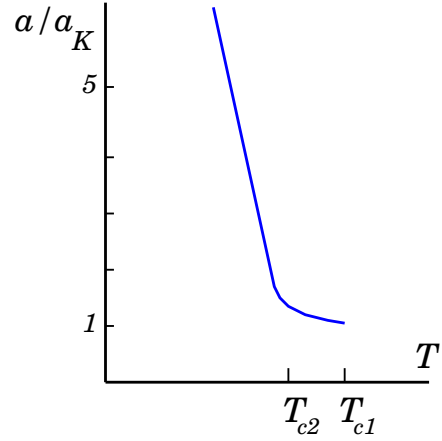}
\caption{ The domain width in slightly inhomogeneous ferroelectric or
ferroelastic in the units of $a_{K}$, the Kittel width (\ref{eq:aKel}). $%
a=a_{K}$ when the domain structure sets in at $T\approx T_{c1}$, and then it
grows linearly with the temperature to large values $a\gg a_{K}$. }
\label{fig:aak}
\end{figure}

Incidentally, close to $T_{c2}$ the domain width is $a\approx \left[ \frac{%
\pi ^{2}\epsilon _{a}^{1/2}\epsilon _{c2}^{1/2}}{14\zeta (3)}\Delta _{1}l_{1}%
\right] ^{1/2}\propto \epsilon _{c2}^{1/4},$ which formally diverges $%
\propto \left( T-T_{c2}\right) ^{-1/4}.$ However, in the vicinity of $T_{c2}$
the induced polarization in the formerly ``hard'' part has about the same
value as the spontaneous polarization in the ``soft'' part, $P_{z2}\approx
P_{01}.$ Then the equation of state in the bottom part becomes strongly
non-linear, since the cubic term is much larger than the linear term, $%
BP_{z2}^{3}\approx BP_{01}^{3}=AP_{01}\approx AP_{z2}\gg A_{2}P_{z2}$, in
the equation of state (since $A\gg A_{2}$ close to $T_{c2})$, so the
response of the bottom layer does not actually soften in this region. In
this case our derivation does not apply, but it is practically certain that
the domain structure in the vicinity of $T_{c2}$ would evolve continuously
upon cooling, Fig. \ref{fig:aak}.

{\em Domain structure at low temperatures } ($T<T_{c2},${\em \ }$A<0,${\em \ 
}$A+\delta A<0).-$ When the system is cooled to below the critical
temperature $T_{c2}$, a spontaneous polarization $\left| P_{02}\right| =%
\sqrt{-A_{2}/B}$ also appears in the bottom layer. The domain structure
simultaneously develops in the whole crystal with domain walls running
parallel to the ferroelectric axis through the whole crystal (if they were
discontinuous at the interface between the two parts of the crystal this
would have created a large depolarizing electric field). The electrostatic
energy requires a solution of the same Laplace equations (\ref{eq:lap1}) and
(\ref{eq:lap2}), only the boundary condition (\ref{eq:bc01})\ would now read 
\begin{equation}
\epsilon _{c1}\varphi _{1k}^{\prime }-\epsilon _{c2}\varphi _{2k}^{\prime
}=4\pi (P_{01k}-P_{02k}),
\end{equation}
where $\epsilon _{c1(2)}=1+2\pi /|A_{1(2)}|\approx 2\pi /|A_{1(2)}|.$ Note
that the density of the bound charge at the interface, corresponding to this
discontinuity of spontaneous polarization, is now $\sigma
_{k}=-(P_{01k}-P_{02k}).$ Therefore, we immediately obtain for the total
free energy of the structure, analogously to the previous case (\ref{eq:Ft01}%
), 
\begin{eqnarray}
\frac{\tilde{F}}{{\cal A}} &=&\frac{P_{01}^{2}\Delta
_{1}l_{1}+P_{02}^{2}\Delta _{2}l_{2}}{a}  \label{eq:Ft12} \\
&&+\frac{16(P_{01}-P_{02})^{2}a}{\pi ^{2}}\sum_{j=0}^{\infty }\frac{1}{%
\left( 2j+1\right) ^{3}D_{2j+1}},
\end{eqnarray}
where $\Delta _{1(2)}=d_{at}\sqrt{|A_{1(2)}|}.$ Not very close to $T_{c2}$
we would have $\sqrt{\frac{\epsilon _{a}}{\epsilon _{c2}}}kl_{2}\gtrsim 1,$
and the minimum of the free energy $\tilde{F}$ is achieved for the domain
width 
\begin{eqnarray}
a &=&\frac{1}{1-P_{02}/P_{01}}  \nonumber \\
&\times &\left[ \frac{\pi ^{2}\epsilon _{a}^{1/2}\left( \epsilon
_{c1}^{1/2}+\epsilon _{c2}^{1/2}\right) }{14\zeta (3)}\left( \Delta
_{1}l_{1}+\Delta _{2}l_{2}\frac{P_{02}^{2}}{P_{01}^{2}}\right) \right]
^{1/2}.
\end{eqnarray}
Close to the critical point $T_{c2}$ the domain width formally behaves as $%
a\propto \epsilon _{c2}^{1/4}\propto (T_{c2}-T)^{-1/4}$, as found just above 
$T_{c2}$ before. The same argument indicates though that our derivation does
not apply in this region, but non-linearity should not cause a substantial
change in the domain structure.

With lowering the temperature to the region where $|A|\gg \delta A,$ we will
have $P_{02}/P_{01}=\sqrt{(A+\delta A)/A}\approx 1+\delta A/2A,$ so that $%
1-P_{02}/P_{01}\approx 2|A|/\delta A\gg 1$ becomes a large prefactor. Note
that in this region $\epsilon _{c1}\approx \epsilon _{c1}=2\pi /|A|,$ $%
\Delta _{1}\approx \Delta _{2}=d_{at}\sqrt{|A|},$ and the domain width
evolves as 
\begin{equation}
a=\frac{|A|}{\delta A}\left[ \frac{2^{5/2}\pi ^{5/2}\epsilon _{a}^{1/2}}{%
7\zeta (3)}d_{at}(l_{1}+l_{2})\right] ^{1/2},
\end{equation}
It becomes much larger than the Kittel width, 
\begin{equation}
\frac{a}{a_{K}}=2^{3/2}\left( \frac{l_{1}+l_{2}}{l_{1}}\right) ^{1/2}\frac{%
T_{c1}-T}{T_{c1}-T_{c2}}\gg 1,
\end{equation}
growing linearly with lowering temperature, if the pinning of the domain
walls is negligible (Fig. \ref{fig:aak}). 
Close to the lower critical point the linearized equation of state does not
apply but the response of the bottom layer remains finite, and we expect, as
mentioned above, that the domain structure would evolve rather gradually
across $T_{c2}$, Fig. \ref{fig:aak}.

Summarizing, in a ferroelectric sample with a tiny inhomogeneity of either
the critical temperature or temperature itself (i.e. in the presence of a
slight temperature gradient and/or minute compositional inhomogeneity across
the system) the domain structure abruptly sets in when the spontaneous
polarization appears in the softest part of the sample (i.e. the part with
maximal $T_{c}$). This takes place when the difference in $T_{c}$ in the
parts of the sample is just $(10^{-3}-10^{-2}){\rm K}$ for displacive
systems, and even smaller, $\left( 10^{-5}-10^{-4}\right) ${\rm K,} for
order-disorder systems. The period of the structure then grows linearly with
lowering temperature and quickly becomes {\em much larger} than the
corresponding Kittel period.

This result does not depend on specific geometry assumed in the present
model. Indeed, if local $T_{c}=T_{c}(z)$ varies continuously, like in graded
ferroelectrics \cite{graded}, it can be approximated by a piece-wise
distribution of a sequence of ``slices''. Upon cooling the system first
looses stability in the softest part of thickness $l_{s},$ which is derived
from the position of the boundary where local $T_{c}=0,$ with respect to a
domain structure with fine period $\propto \sqrt{l_{s}}.$ The domains extend
into the bulk of the system and become wider with further cooling, since $%
l_{s}$ increases. In electroded sample there will be no branching and domain
walls would run straight across all transformed slices. Otherwise,
discontinuities would have resulted in very strong depolarizing field. If
the overall inhomogeneity is small, the picture would obviously remain very
similar to the two-slice model solved above. The same arguments remain valid
if the inhomogeneity were to have more complex form/distribution in a
sample. The novel feature of the present effect of the depolarizing field is
that it appears not due to surface charges, which are screened out by the
electrodes, but because of the bulk inhomogeneity. The bulk depolarizing 
fields are present in other important classes of inhomogeneous
ferroelectrics, {\em graded} ferroelectric films\cite{graded} and
superlattices of different ferroelectrics (e.g. KNbO$_{3}$/KTaO$_{3})$\cite
{FEsuper}, and may be responsible for their unusual behavior.

We have shown that a very tiny temperature gradient, or a slight
compositional inhomogeneity, etc., would result in practically any crystal
eventually splitting into domains no matter how high the quality of it is.
The unusual evolution of the domain pattern, found in the present paper,
when it starts from very fine domains at $T_{c}$ and then grows linearly
with temperature to very large sizes, has been reported in Ref. \cite
{nakatani85} for $\sim 1$mm thick TGS\ crystals. It is worth noting that the
result is very general and applies also to slightly inhomogeneous free
ferroelastic crystals\cite{BLinhelast}. Other implications include
extensively studied graded films and ferroelectric superlattices\cite
{graded,FEsuper}. It would be very interesting to perform controlled
experiments for the domain structure close to the phase transition to check
the present theory.

\end{document}